\documentclass{icrc29_astro}
\usepackage{graphicx,amssymb,amsmath,times}
\newcommand{\SC}{self-consistent}
\newcommand{\delgam}{\delta\theta \ll 1/\gamZ}
\newcommand{\sigKW}{\sigma_\mathrm{KW}}
\newcommand{\PLn}{power law}
\newcommand{\PL}{power-law}

\newcommand{\deltime}{\delta t}

\newcommand{\etamfp}{\eta_\mathrm{mfp}}
\newcommand{\MC}{Monte Carlo}
\newcommand{\mc}{Monte Carlo}
\newcommand{\Lor}{Lorentz}
\newcommand{\fofp}{f(p)}
\newcommand{\degg}{^\circ}
\newcommand{\Tbn}{\theta_\mathrm{B0}}

\newcommand{\gamZ}{\gamma_0}

\newcommand{\NL}{nonlinear}
\newcommand{\GRB}{$\gamma$-ray burst}
\newcommand{\GRBs}{$\gamma$-ray bursts}
\newcommand{\rel}{relativistic}
\newcommand{\Rel}{Relativistic}
\newcommand{\nonrel}{non\-rel\-a\-tiv\-is\-tic}
\newcommand{\transrel}{trans-rel\-a\-tiv\-is\-tic}
\newcommand{\ultrarel}{ul\-tra-rel\-a\-tiv\-is\-tic}
\newcommand{\TP}{test-particle}

\newcount\listno
\def\List{\global\advance \listno by 1 {(\the\listno)}}

\newcount\listcno
\def\Listc{\global\advance \listcno by 1 
	{({\expandafter{\romannumeral\listcno})\,}}}
	\def\newlistc{\listcno=0}
\begin{document}
\title[DSA in \Rel\ Shocks]{Test-Particle Diffusive Shock
  Acceleration in Relativistic Shocks}
\author[Donald C. Ellison] {Donald C. Ellison\\
        Physics Dept., North Carolina State Univ., Raleigh, NC 27695-8202, U.S.A.
        }
\presenter{Presenter: Don Ellison (don$\_$ellison@ncsu.edu), \  
usa-ellison-D-abs2-og24-poster}

\maketitle

\begin{abstract}
We present results from a fully \rel\ Monte Carlo simulation of diffusive shock
acceleration (DSA) in unmodified (i.e., \TP) shocks. The computer code uses a single
algorithmic sequence to smoothly span the range from \nonrel\ speeds to fully \rel\
shocks of arbitrary obliquity, providing a powerful consistency check.
We show the dependence of the particle spectrum, which can differ strongly from the
canonical $\fofp \propto p^{-4.23}$ result, on the obliquity and on the strength and
``fineness'' of scattering for a range of shock \Lor\ factors. The \MC\ results are
also shown to be consistent with a simple relation for the spectral index given by
Keshet and Waxman\cite{KW05} in parallel shocks when the diffusion is ``fine'' and
isotropic.
\end{abstract}

\vskip-24pt\hbox{}
\section{Introduction} \vskip-6pt
Most work on diffusive shock acceleration (DSA) in \rel\ shocks has been restricted
to the \ultrarel\ regime with magnetic fields assumed parallel to the shock
normal. However, \transrel\ shocks are certain to be important in some sources and,
in general, \rel\ shocks will have highly oblique magnetic fields.
Here we consider \TP\ diffusive shock acceleration in shocks of arbitrary
obliquity\footnote{Oblique shocks are those where the angle between the upstream
magnetic field and the shock normal, $\Tbn$, is greater than $0\degg$. Parallel
shocks are those with $\Tbn=0\degg$. Everywhere in this paper we use the subscript 0
(2) to indicate upstream (downstream) quantities.}
and with arbitrary Lorentz factors, $\gamZ$.
We consider only particle acceleration in plane, unmodified (i.e., \TP) shocks where
effects on the shock structure from superthermal particles are ignored.  In contrast
to \nonrel\ shocks, the details of particle scattering strongly influence the
superthermal particle populations produced in \transrel\ and \ultrarel\ shocks (see
\cite{BedOstrow96}\cite{BedOstrow98}\cite{NO04} for example).  Unfortunately, these
details are not known with any reliability so the results of all current models of
DSA depend on the particular scattering assumptions made.
We use a simple, parameterized model of particle diffusion (described in detail in
\cite{ED04}) which we believe adequately describes the main features of particle
transport. Until a self-consistent theory of wave-particle interactions in \rel\
shocks is produced, parameterization will be necessary.

\vskip-24pt\hbox{}
\section{Model and Results} \vskip-6pt
The acceleration process is modeled with a \MC\ simulation where thermal particles
are injected far upstream and followed as they diffuse through the shock (for full
details, see \cite{ED04} and references therein). Scattering is elastic and isotropic
in the local plasma frame and the scattering mean free path $\lambda$ in the local
frame is proportional to the gyroradius $r_g$.  We simulate small-angle scattering by
allowing the tip of the particle's fluid-frame momentum vector $\mathbf{p}$ to
undergo a random walk on the surface of a sphere.  After a small time ${\delta t}$
the momentum undergoes a small change in direction of magnitude ${\delta\theta}$
within a maximum angle $\delta\theta_{\rm max}$.

If the time in the local frame required to accumulate deflections of the order of
$90^\circ$ is identified with the collision time $t_c = \lambda /v_p$, \cite{EJR90}
showed that $\delta \theta_{\rm max} = \sqrt{ 6 \delta t /t_c }$,
%
%
where  $\delta t$ is the time between pitch-angle scatterings.
We take $\lambda$ proportional to the gyroradius $r_g = pc/(eB)$ ($e$ is the
electronic charge and $B$ is the local uniform magnetic field in Gaussian units),
i.e., $\lambda = \etamfp \, r_g$, where $\etamfp$ is a measure of the ``strength'' of
scattering. The strong scattering limit, $\etamfp=1$, is called Bohm diffusion and in
this limit, cross-field diffusion is important.
Setting $\deltime = \tau_g / N_g$, where $N_g$ is the number of gyro-time
segments $\deltime$, dividing a gyro-period $\tau_g = 2 \pi r_g/v_p$, we have
$\delta \theta_{\rm max} = \sqrt{12 \pi / (\etamfp N_g)}$.
%
%
Large values of $N_g$ result in ``fine'' scattering where it is assumed that magnetic
field fluctuations exist on all scales so that ${\delta\theta}$ can be arbitrarily
small.
The DSA distribution, $\fofp$, becomes independent of $N_g$ when it is large enough
such that ${\delta\theta} \ll 1/\gamZ$, otherwise large-angle scattering (LAS)
effects become important.
Thus, the scattering properties of the medium are modeled with the two parameters
$\etamfp$ and $N_g$ and, along with $\gamZ$ and $\Tbn$, this yields, in the high Mach
number limit, four independent parameters characterizing DSA.

\begin{figure}[h]
\begin{center}
\includegraphics*[width=0.7\textwidth,angle=0,clip]{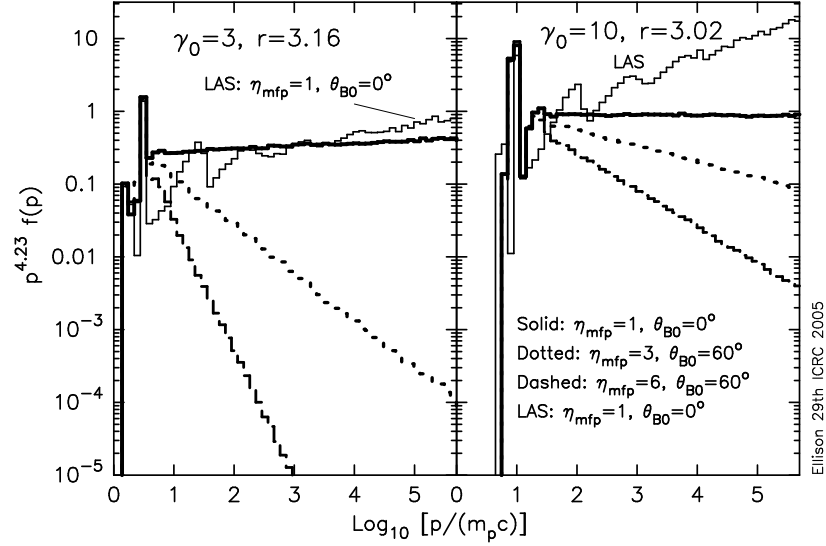}
\caption{\label {fig1} Test-particle spectra for shock \Lor\ factors $\gamZ=3$ (left
  panel) and $\gamZ=10$ (right panel) (note that $p^{4.23} \fofp$ is plotted). For
  $\gamZ=3$, the shock compression ratio, as determined in \cite{DBJE2004}, is
  $r=3.16$. For $\gamZ=10$, $r=3.02$. The listing of line types applies to both
  panels. The light-weight solid curves labeled LAS show the effects of large-angle
  scattering, all other curves have ${\delta\theta} \ll 1/\gamZ$. All spectra are
  calculated at the shock in the shock frame.}
\end{center}
\end{figure}

In Fig.~\ref{fig1} we show examples where these four parameters are varied. We plot
$p^{4.23} \fofp$ for a \transrel\ $\gamZ=3$ and a more fully \rel\ $\gamZ=10$. In all
cases except the two marked LAS, $N_g$ is large enough so that ${\delta\theta} \ll
1/\gamZ$. In the two LAS cases, the effects
of large-angle scattering are clearly seen.

In the limits ${\delta\theta} \ll 1/\gamZ$ and $\gamZ \gg 1$, DSA produces a power
law $\fofp \propto p^{-4.23}$, independent of $\Tbn$ or $\etamfp$ (see
\cite{BedOstrow98}).
The heavy solid
curve in the right panel of Fig.~\ref{fig1} shows that for $\Tbn=0\degg$ and
$\etamfp=1$, the canonical \PLn\ is obtained. However, even with ${\delta\theta} \ll
1/\gamZ$, $\fofp$ can be much steeper than this for larger values of $\Tbn$ and
$\etamfp$. We showed in \cite{ED04} that effects of $\Tbn$ and $\etamfp$ diminish with
increasing $\gamZ$ but persist noticeably to at least $\gamZ = 30$.

The curves labeled LAS in Fig.~\ref{fig1} both have $\Tbn=0\degg$ and $\etamfp=1$, but
here $N_g$ is small enough so particles make large deflections in individual
scattering events. 
Large-angle scattering will be important whenever $\etamfp N_g$ is small enough so
that $\delta\theta_{\rm max} \not\ll 1/\gamZ$ and, in principle, could occur even in
the shocks expected when a \GRB\ fireball expands \ultrarel ally into the
interstellar environment.

\begin{figure}[h]
\begin{center}
\includegraphics*[width=0.6\textwidth,angle=0,clip]{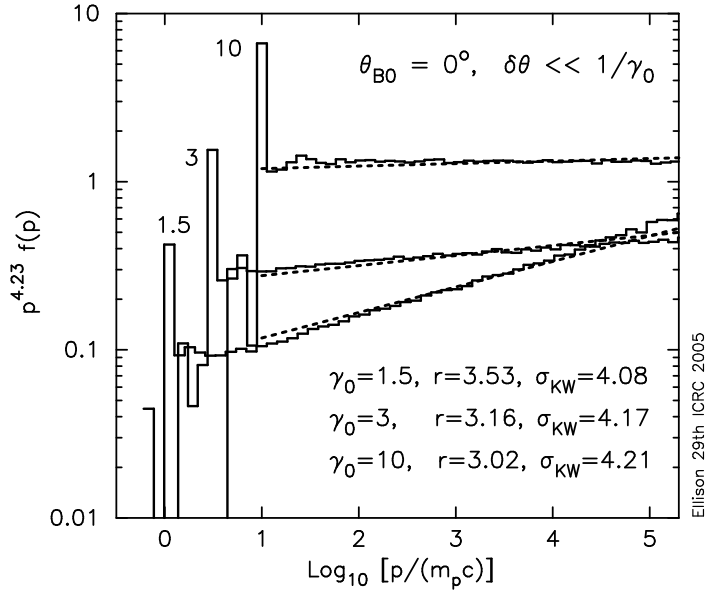}
\caption{\label {fig2} Comparison of \mc\ spectra (solid histograms) with the
  analytic result given in \cite{KW05} (dotted lines). $p^{4.23}\fofp$ is plotted and
  the normalization is arbitrary.}
\end{center}
\end{figure}

Keshet and Waxman\cite{KW05} have presented the following analytic expression for the
\PL\ spectral index $\sigKW$
\begin{equation}
\label{eq:Wax}
\sigKW =\left ({3\beta_0 - 2\beta_0 \beta_2^2 +\beta_2^3}\right) /
\left({\beta_0 - \beta_2}\right)
\ ,
\end{equation}
where $\beta_0$ ($\beta_2$) is the speed of the upstream (downstream) fluid
normalized to the speed of light. This expression, which assumes that the shock is
parallel and that $\delgam$, depends only on the shock \Lor\ factor and the
compression ratio. The compression ratio must be determined independently. In
Fig.~\ref{fig2} we compare \mc\ spectra to eq.~(\ref{eq:Wax}) for several $\gamZ$.
The \mc\ results use the solution to the shock jump conditions (including the
equation of state) calculated in \cite{DBJE2004} to determine $r$ and this value is
used in eq.~(\ref{eq:Wax}) (i.e., $\beta_2 = \beta_0/r$) and listed in
Fig.~\ref{fig2}.  The match between the two models is quite good in this \transrel\
range and will improve outside of this range.

\vskip-24pt\hbox{}
\section{Discussion and Conclusions} \vskip-6pt
In \nonrel\ shocks, the \TP\ power law from DSA depends only on the shock
compression; it is independent of the details of diffusion, on the strength of
scattering, and on the obliquity (i.e., independent of $N_g$, $\etamfp$, and $\Tbn$).
In \rel\ shocks, particularly \transrel\ ones, \TP\ spectra can depend strongly on
these parameters.
We have shown examples with various $N_g$, $\etamfp$, and $\Tbn$ and, in general, the
\PLn\ steepens if $\etamfp > 1$ and/or $\Tbn > 0\degg$. The parameter $N_g$
determines the fineness of scattering and if $N_g$ is small enough so that
$\delta\theta_{\rm max} \not\ll 1/\gamZ$,
large departures
from a \PLn\ can occur and the spectrum can become harder  than $p^{-4.23}$ even for
\ultrarel\ shocks.
In addition, as have shown that the simple analytic result given in
eq.~(\ref{eq:Wax}) \cite{KW05} matches our \mc\ results for parallel shocks and when
$\delgam$.

\newlistc

The parameterization of diffusion used in the \mc\ technique is useful for
investigating the behavior of DSA in \rel\ shocks. It makes fewer approximations than
analytic models and can cover a far larger dynamic range than any current
particle-in-cell (PIC) simulation.  Nevertheless, the actual diffusion depends on the
\SC\ generation of magnetic turbulence by the accelerated particles and is 
inherently complex and has not been described adequately even for \nonrel\ shocks
(see, for example, \cite{MD01}). 
In fact, the correct form for the magnetic power spectrum in self-generated, \rel\
turbulence will not be known until PIC
simulations can perform this calculation. Unfortunately,
the treatment of these problems using PIC codes is beyond current
computing capabilities for three basic reasons:
\Listc\ 
To correctly describe cross-field diffusion and other physics of the viscous
subshock, PIC simulations must be done fully in 3-D \cite{JKG93}\cite{JJB98}. If 1-
or 2-D simulations are used, cross-field diffusion, an essential element in DSA, is
unrealistically suppressed;
\Listc\ 
If electrons are to be understood, electrons and protons must be modeled
simultaneously (except for pair plasma shocks), and the simulation must treat widely
disparate inertial scales, greatly adding to the run time; and
\Listc\ 
In order for nonlinear effects to become apparent or field
generation to occur on large scales, simulations must be run long enough in a large
enough box with enough particles for a significant population of extremely energetic
particles to be produced.
The combination of these three requirements means that, while PIC simulations can add
to our knowledge of critical aspects of the injection problem and the start of
magnetic field generation process (e.g., \cite{Nish05}), they will not be able to
model the injection and \NL\ acceleration of electrons and ions to energies relevant
to ultra-high energy cosmic rays or GRBs until computers more powerful than exist
today become available.

While we have only discussed \TP\ acceleration, \NL\ effects, where the accelerated
particles modify the shock structure, may be important in \rel\ shocks (see
\cite{ED02}), and in fact, most models of \GRBs\ assume that \rel\ shocks efficiently
accelerate electrons.  If this is the case, \TP\ spectra may not be good
approximations. This is certainly true for DSA in \transrel\ shocks, where the harder
intrinsic \PLn\ will naturally lead to a \NL\ feedback on the shock structure.

Nonlinear effects will also influence the distribution of energy between electrons
and protons (e.g., \cite{Double03}) and modify the shape of the distribution at low energies,
i.e., near the sharp peaks at $\sim \gamZ$ in Figs.~\ref{fig1} and \ref{fig2} (i.e.,
\cite{ED02}). These issues are critical for \GRB\ models.

{\bf Acknowledgements:} D.C.E. wishes to acknowledge support from a NASA grant
(ATP02-0042-0006) and the KITP (Santa Barbara) under NSF Grant No. PHY99-0794.

\newcommand\itt{ }
\newcommand\bff{ }
\newcommand{\aaDE}[3]{ 19#1, A\&A, #2, #3}
\newcommand{\aatwoDE}[3]{ 20#1, A\&A, #2, #3}
\newcommand{\aatwopress}[1]{ 20#1, A\&A, in press}
\newcommand{\aasupDE}[3]{ 19#1, {\itt A\&AS,} {\bff #2}, #3}
\newcommand{\ajDE}[3]{ 19#1, {\itt AJ,} {\bff #2}, #3}
\newcommand{\anngeophysDE}[3]{ 19#1, {\itt Ann. Geophys.,} {\bff #2}, #3}
\newcommand{\anngeophysicDE}[3]{ 19#1, {\itt Ann. Geophysicae,} {\bff #2}, #3}
\newcommand{\annrevDE}[3]{ 19#1, {\itt Ann. Rev. Astr. Ap.,} {\bff #2}, #3}
\newcommand{\apjDE}[3]{ 19#1, {\itt ApJ,} {\bff #2}, #3}
\newcommand{\apjtwoDE}[3]{ 20#1, {\itt ApJ,} {\bff #2}, #3}
\newcommand{\apjletDE}[3]{ 19#1, {\itt ApJ,} {\bff  #2}, #3}
\newcommand{\apjlettwoDE}[3]{ 20#1, {\itt ApJ,} {\bff  #2}, #3}
\newcommand{\apjpress}{{\itt ApJ,} in press}
\newcommand{\apjletpress}{{\itt ApJ(Letts),} in press}
\newcommand{\apjsDE}[3]{ 19#1, {\itt ApJS,} {\bff #2}, #3}
\newcommand{\apjstwoDE}[3]{ 19#1, {\itt ApJS,} {\bff #2}, #3}
\newcommand{\apjsubDE}[1]{ 19#1, {\itt ApJ}, submitted.}
\newcommand{\apjsubtwoDE}[1]{ 20#1, {\itt ApJ}, submitted.}
\newcommand{\appDE}[3]{ 19#1, {\itt Astropart. Phys.,} {\bff #2}, #3}
\newcommand{\apptwoDE}[3]{ 20#1, {\itt Astropart. Phys.,} {\bff #2}, #3}
\newcommand{\araaDE}[3]{ 19#1, {\itt ARA\&A,} {\bff #2},
   #3}
\newcommand{\assDE}[3]{ 19#1, {\itt Astr. Sp. Sci.,} {\bff #2}, #3}
\newcommand{\grlDE}[3]{ 19#1, {\itt G.R.L., } {\bff #2}, #3} 
\newcommand{\icrcplovdiv}[2]{ 1977, in {\itt Proc. 15th ICRC (Plovdiv)},
   {\bff #1}, #2}
\newcommand{\icrcsaltlake}[2]{ 1999, {\itt Proc. 26th Int. Cosmic Ray Conf.
    (Salt Lake City),} {\bff #1}, #2}
\newcommand{\icrcsaltlakepress}[2]{ 19#1, {\itt Proc. 26th Int. Cosmic Ray Conf.
    (Salt Lake City),} paper #2}
\newcommand{\icrchamburg}[2]{ 2001, {\itt Proc. 27th Int. Cosmic Ray Conf.
    (Hamburg),} {\bff #1}, #2}
\newcommand{\JETPDE}[3]{ 19#1, {\itt JETP, } {\bff #2}, #3}
\newcommand{\jgrDE}[3]{ 19#1, {\itt J.G.R., } {\bff #2}, #3}
\newcommand{\mnrasDE}[3]{ 19#1, {\itt MNRAS,} {\bff #2}, #3}
\newcommand{\mnrastwoDE}[3]{ 20#1, {\itt MNRAS,} {\bff #2}, #3}
\newcommand{\mnraspress}[1]{ 20#1, {\itt MNRAS,} in press}
\newcommand{\natureDE}[3]{ 19#1, {\itt Nature,} {\bff #2}, #3}
\newcommand{\naturetwoDE}[3]{ 20#1, {\itt Nature,} {\bff #2}, #3}
\newcommand{\nucphysA}[3]{#1, {\itt Nuclear Phys. A,} {\bff #2}, #3}
\newcommand{\pfDE}[3]{ 19#1, {\itt Phys. Fluids,} {\bff #2}, #3}
\newcommand{\phyreptsDE}[3]{ 19#1, {\itt Phys. Repts.,} {\bff #2}, #3}
\newcommand{\physrevEDE}[3]{ 19#1, {\itt Phys. Rev. E,} {\bff #2}, #3}
\newcommand{\prlDE}[3]{ 19#1, {\itt Phys. Rev. Letts,} {\bff #2}, #3}
\newcommand{\prltwoDE}[3]{ 20#1, {\itt Phys. Rev. Letts,} {\bff #2}, #3}
\newcommand{\revgeospphyDE}[3]{ 19#1, {\itt Rev. Geophys and Sp. Phys.,}
   {\bff #2}, #3}
\newcommand{\rppDE}[3]{ 19#1, {\itt Rep. Prog. Phys.,} {\bff #2}, #3}
\newcommand{\rpptwoDE}[3]{ 20#1, {\itt Rep. Prog. Phys.,} {\bff #2}, #3}
\newcommand{\ssrDE}[3]{ 19#1, {\itt Space Sci. Rev.,} {\bff #2}, #3}
\newcommand{\ssrtwoDE}[3]{ 20#1, {\itt Space Sci. Rev.,} {\bff #2}, #3}
\newcommand{\scienceDE}[3]{ 19#1, {\itt Science,} {\bff #2}, #3} 
\newcommand{\spDE}[3]{ 19#1, {\itt Solar Phys.,} {\bff #2}, #3} 
\newcommand{\spuDE}[3]{ 19#1, {\itt Sov. Phys. Usp.,} {\bff #2}, #3}

\end{document}